\documentclass[
   ,draft            
  ,numberedheadings 
  ]
  {aipproc}
\usepackage{epsfig}
\layoutstyle{6x9}

\begin{document}

\newcommand{\be}{\begin{equation}}
\newcommand{\ee}{\end{equation}}
\newcommand{\bea}{\begin{eqnarray}}
\newcommand{\eea}{\end{eqnarray}}
\newcommand{\da}{\dagger}
\newcommand{\dg}[1]{\mbox{${#1}^{\dagger}$}}
\newcommand{\hlf}{\mbox{$1\over2$}}
\newcommand{\lfrac}[2]{\mbox{${#1}\over{#2}$}}
\newcommand{\scsz}[1]{\mbox{\scriptsize ${#1}$}}
\newcommand{\tsz}[1]{\mbox{\tiny ${#1}$}}
\newcommand{\ch}       {\mbox{\bf (*** CHECK! ***)~}}
\renewcommand{\descriptionlabel}[1]%
		{\hspace{\labelsep}\textsf{#1}}

\title{The Pioneer Anomaly: \\
The Data, its Meaning, and a Future Test}

\author{$^1$\footnote{Speaker: M.M.N. 
Email addresses:  mmn@lanl.gov, turyshev@jpl.nasa.gov, \newline
john.d.anderson@jpl.nasa.gov}Michael Martin Nieto}{
  address={Theoretical Division (MS-B285), Los Alamos National Laboratory,\\
University of California,  Los Alamos, New Mexico 87545}
}

\author{Slava G. Turyshev}{
  address={Jet Propulsion Laboratory, California Institute of  Technology,\\
Pasadena, CA 91109, U.S.A.}
}

\author{John D. Anderson}{
  address={Jet Propulsion Laboratory, California Institute of  Technology,\\
Pasadena, CA 91109, U.S.A.}
}

\begin{abstract}
The radio-metric Doppler tracking data from the Pioneer 10/11
spacecraft, from between 20-70 AU, yields an unambiguous and
independently confirmed anomalous  blue shift drift of 
$a_t = (2.92 \pm 0.44)\times 10^{-18}$ s/s$^2$.  It   
can be interpreted as being 
due to a constant acceleration of $a_P = (8.74 \pm 1.33) \times
10^{-8}$  cm/s$^2$ directed towards the Sun.  No systematic effect has
been able to explain the anomaly, even though such an origin is an
obvious candidate.  We discuss what has been learned 
(and what might still be learned) from the data about the anomaly, its 
origin, and the mission design characteristics that would be needed to 
test it.  Future mission options are proposed.
\end{abstract}

\maketitle



\section{The Data} 


\subsection{The Pioneer missions and the anomaly}
\label{orig}

\begin{figure}[ht]
\psfig{figure=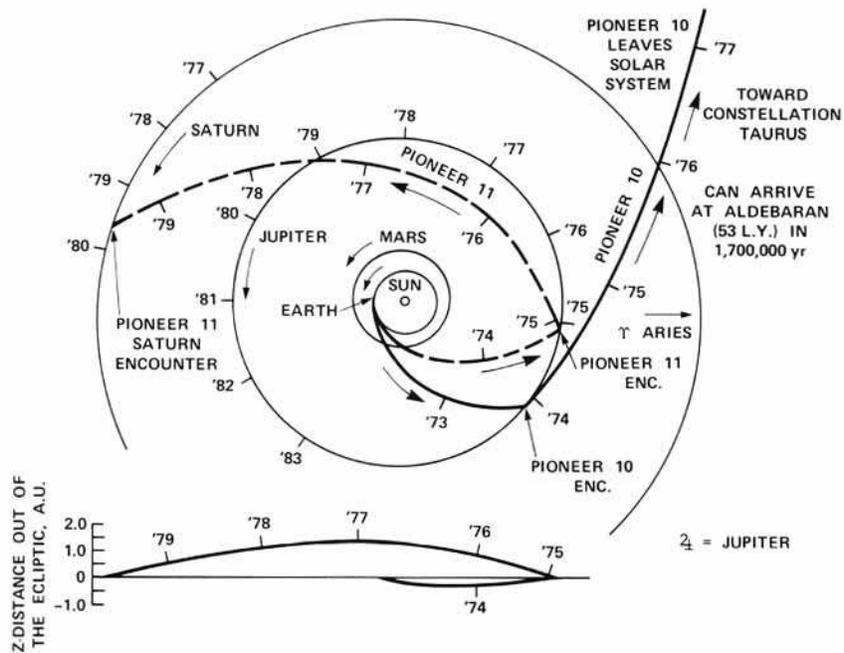,width=4.5in}
\caption
{The early trajectories of the Pioneer 10 and 11 spacecraft, out to the orbit of Saturn. 
\label{pioclosepath}}
\end{figure}

The Pioneer 10/11 missions, launched on 2 March 1972 (Pioneer 10) and 4 December 1973 (Pioneer 11), were the first to explore the outer solar system.
After Jupiter and (for Pioneer 11) Saturn encounters, the two 
spacecraft followed escape hyperbolic orbits near the plane of the 
ecliptic to opposite sides of the solar system. (See Figure
\ref{pioclosepath}.) Pioneer 10 eventually 
became the first man-made object to leave the solar system.

The Pioneers were excellent crafts with which to perform precise 
celestial mechanics experiments. This was due to a combination of many 
factors, including (1) their attitude control -- spin-stabilized, with a 
minimum number of commanded attitude correction maneuvers using 
thrusters, (2) power design -- Radioisotope Thermoelectric Generators 
(RTGs) on extended booms aiding the stability of the craft and also 
reducing the heat systematics, and (3) precise Doppler tracking -- with 
sensitivity to resolve small frequency drifts at the level of mHz/s). 
The result was the most precise navigation in deep space to date.


\begin{figure}[h!]
\psfig{figure=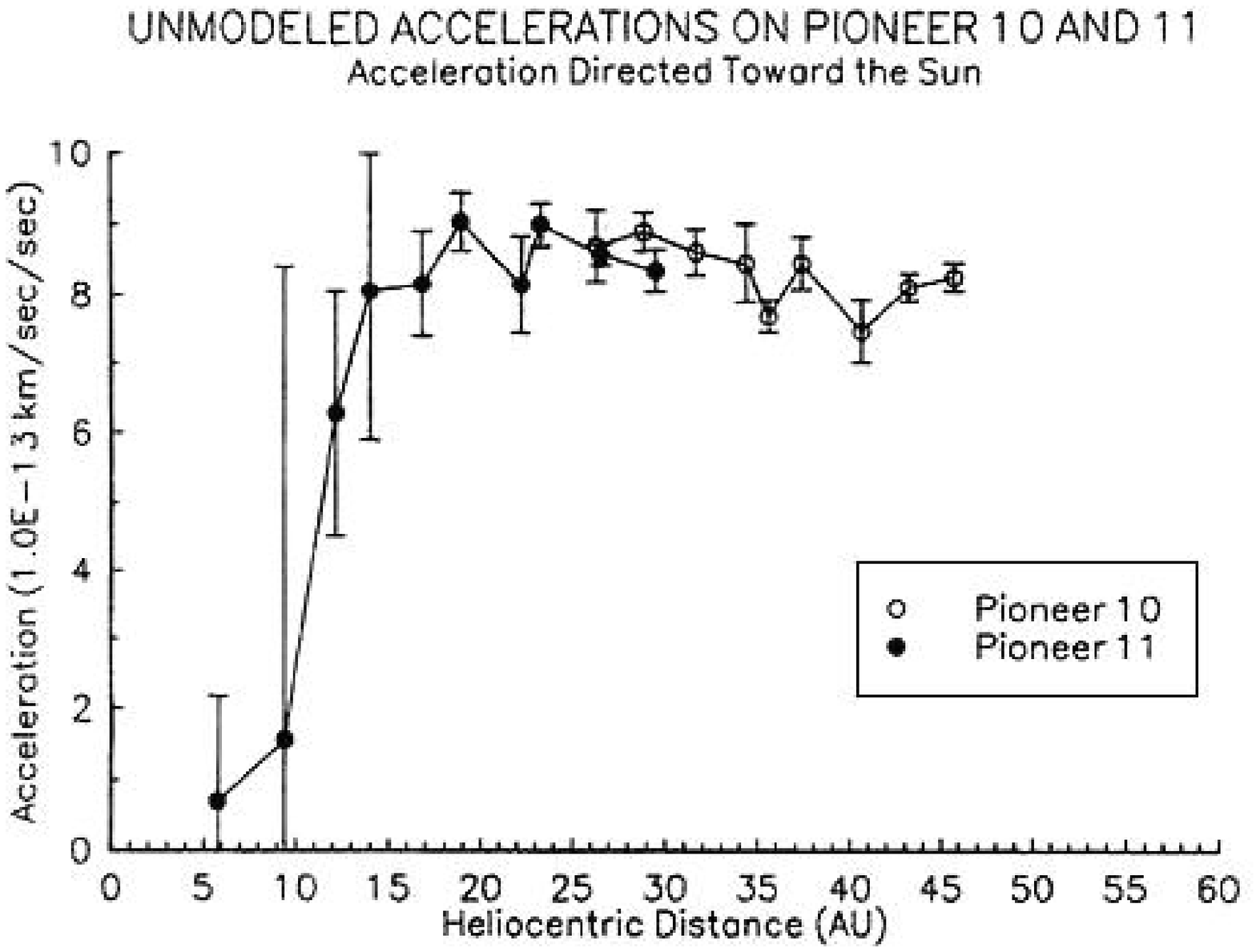,width=4.5in}
\caption
{A plot of the early unmodeled 
accelerations of Pioneer 10 and Pioneer 11, from about 
1981 to 1989 and 1977 to 1989, respectively \cite{pioprd}.
\label{fig:correlation}}
\end{figure}


By 1980 Pioneer 10 had passed a distance of $\sim$20 AU (Astronomical Units) from the Sun and the 
acceleration contribution from solar-radiation pressure on the craft 
(directed away from the Sun) had decreased to less than $4 \times
10^{-8}$ cm/s$^2$.  
At that time an anomaly in the Doppler signal became evident. 
Therefore, from time to time various pieces of the early data were monitored 
for residuals by different analysts who used 
different criteria for judgments.  But  there never was a detailed look at the
systematics nor a comprehensive look at the whole. 

However, as shown
in Figure \ref{fig:correlation}, these analyses strongly 
indicated an anomaly past 10 AU.   (All the data averages were of order 
$8  \times 10^{-8}$ cm/s$^2$, 
with statistical errors that decreased with distance and were less than $2  \times 10^{-8}$ cm/s$^2$.)

Subsequently, the Pioneer 
Collaboration was formed among the present authors and Philip
Laing, Eunice Lau, and Tony Liu to perform a NASA-sponsored 
analysis of the radio-metric tracking data from the Pioneer 10/11 
spacecraft.   To make sure that the radiation pressure was not an
overwhelming systematic, the data from 1987.0 was considered.  This
covered  distances between 20 - 70 AU from the 
Sun.  (The inner part from Pioneer 11, the outer part from Pioneer 10.)

\begin{figure}[h!]
\epsfig{figure=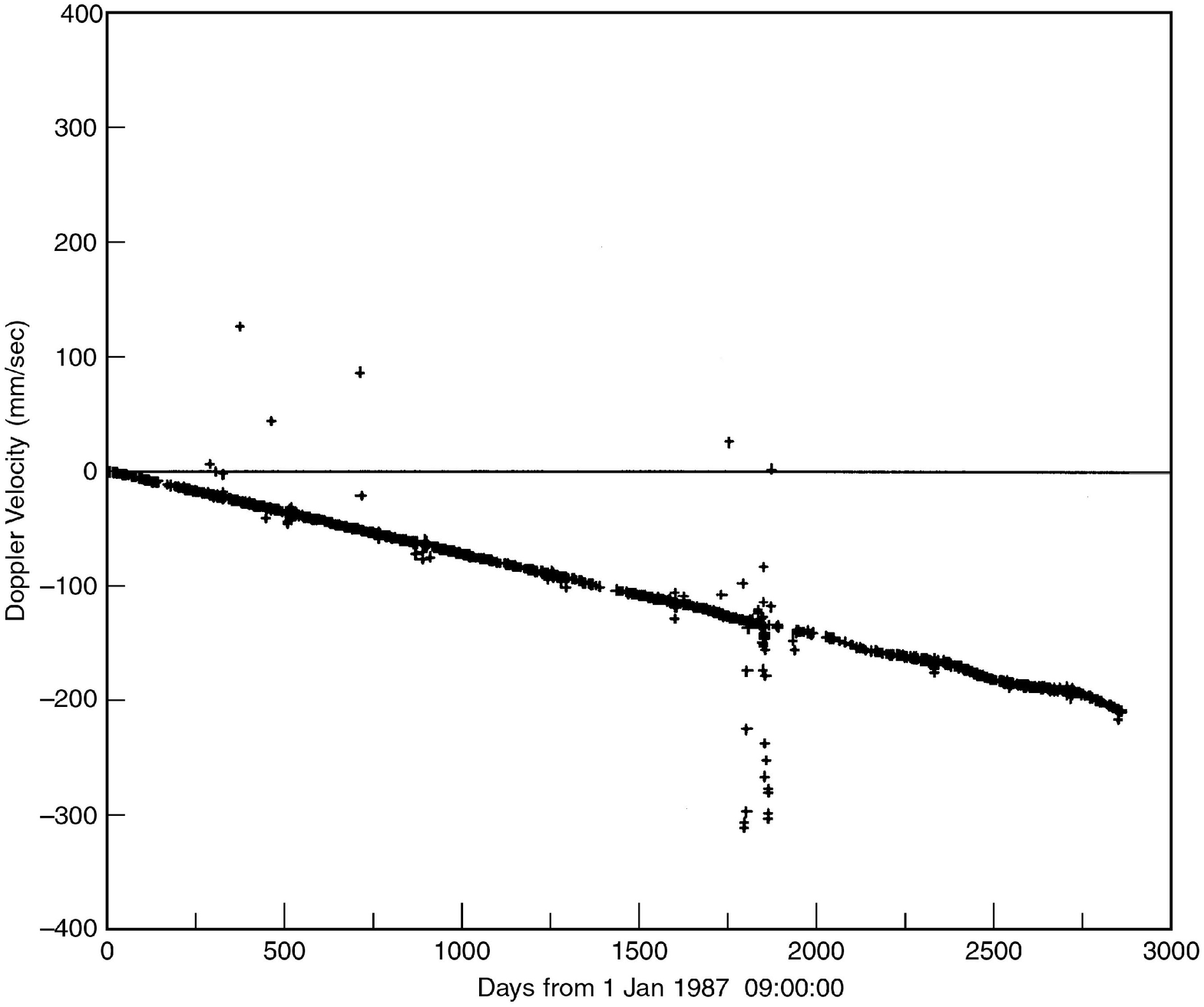,width=4in}
   \caption
{CHASMP best fit for the Pioneer 10 Doppler residuals with the 
anomalous acceleration taken out \cite{pioprl,pioprd}.  
After adding one more parameter to the
model (a constant radial acceleration) the residuals are distributed 
about zero Doppler velocity with a systematic variation $\sim$ 3.0 mm/s
on a time scale of $\sim$ 3 months.  The outliers on the plot were
rejected from the fit. [The quality of the fit may be determined by the
ratio of residuals to the downlink carrier frequency, $\nu_0\approx$
2.29 GHz.]
 \label{chasmp}}
\end{figure}

This data consistently indicated the presence of an anomalous, small, 
constant Doppler frequency drift. In figure \ref{chasmp} we show Pioneer 10 Doppler residuals using The Aerospace Corporation's CHASMP navigation code.


Although the most obvious explanation would be that there is a 
systematic origin to the effect, perhaps generated by the spacecraft 
themselves from excessive heat or propulsion gas leaks, none has been 
found; that is, no unambiguous, onboard systematic has been discovered 
\cite{pioprd,piompla}.   The largest systematics were indeed 
from on-board the craft but, as recounted in Table \ref{error_budget}, they
did not come near to explaining the anomaly \cite{pioprd,piompla}.


\begin{table}[h]
\caption{Large (on-board) Systematics (Biases and Uncertainties) \cite{pioprd}.
\label{error_budget}} 
\begin{tabular}{lll} 
\hline\hline
 Description of error budget constituents & 
Bias~~~~~& Uncertainty    \\     
&        $10^{-8} ~\rm cm/s^2$ &  $10^{-8} ~\rm cm/s^2$  \\\hline
                 &             &         \\

 a) Radio beam reaction force         & $+1.10$&$\pm 0.11$ \\
 b) RTG heat reflected off the craft      &  $-0.55$&$\pm 0.55$ \\
 c) Differential emissivity of the RTGs         & &  $\pm 0.85$ \\
 d) Non-isotropic radiative cooling of the spacecraft && $\pm 0.48$\\
 e) Expelled Helium  produced  within the RTGs 
           &$+0.15$ & $\pm 0.16$   \\
 f) Gas leakage           &         &  $\pm 0.56$   \\
[10pt]
\hline\hline
\end{tabular} 
\end{table}


After taking into account all systematics, 
a blue shift in $(\Delta \nu/\nu)$ was determined of  
$a_t = (2.92 \pm 0.44)\times 10^{-18}$ s/s$^2$.  It  
can be interpreted as being due to  
a constant acceleration of $a_P = (8.74 \pm 1.33) \times
10^{-8}$ cm/s$^2$ directed towards the Sun \cite{pioprl}-\cite{mark}.

Attempts to find a convincing systematic explanation 
have not succeeded \cite{pioprd,piompla}. 
This inability to explain the anomalous 
acceleration of the Pioneer spacecraft with conventional physics has 
contributed to the growing discussion about its origin.

Attempts to verify the anomaly using other spacecraft have proven 
disappointing \cite{pioprl,pioprd}. 
This is because the Voyager, Galileo, Ulysses, 
and Cassini spacecraft navigation data all have their own individual 
difficulties for use in an independent test of the anomaly 
(see below). In addition, 
many of the deep space missions that are currently being planned either 
will not provide the needed navigational accuracy and trajectory 
stability of under $10^{-8}$ cm/s$^2$ (i.e., Pluto Express, InterStellar Probe) 
or else they will have significant on-board systematics that mask the 
anomaly (i.e., JIMO -- Jupiter Icy Moons Orbiter).

The acceleration regime in which the anomaly was observed diminishes the 
value of using modern disturbance compensation systems for a test. For 
example, the systems that are currently being developed for the LISA and 
LISA Pathfinder missions, are designed to operate in the presence of a 
very low frequency acceleration noise (at the mHz level), while the 
Pioneer anomalous acceleration is a strong constant bias in the Doppler 
frequency data. Further, currently available DC (constant) accelerometers are a 
few orders of magnitude less sensitive than is needed for a test. 
Also, should the anomaly not really be a force but rather an effect that universally 
affects frequency standards [2], the use of accelerometers will shed no 
light on what is the true nature of the observed anomaly.


\section{The Data's Meaning}
\subsection{What the data tells us}

The major study \cite{pioprd} demonstrated that:
 
{\it For Pioneer 10,} between about 40 and 70.5 AU (data taken
between 1987.0 and 1998.5), there was an experimental signal of 
\be
a_{P(expt)}^{Pio10}= (7.84 \pm 0.01) \times 10^{-8}~\mathrm{cm/s}^2. 
\ee 

{\it For Pioneer 11,} between about 22.4 and 71.7 AU (data taken
between 1987.0 and 1990.8), there was an experimental signal of 
\be
a_{P(expt)}^{Pio10}= (8.55 \pm 0.02) \times 10^{-8}~\mathrm{cm/s}^2. 
\ee
As stated above, the analysis for both Pioneers with all systematics
included yielded
\be
a_{P}= (8.74 \pm 1.33) \times 10^{-8}~\mathrm{cm/s}^2.
\ee

The effect was seen only on these small ($\sim$ 250 kg) craft on hyperbolic
orbits.  There were indicative signals from the Galileo and Ulysses
craft, but the data was not conclusive.\footnote{
Galileo was only spin-stabilized during Earth-Jupiter cruise.  Although
this data set was useful to verify the Deep Space Network hardware, it
came from so close in to the Sun that it was too highly correlated with
the solar radiation pressure to yield a positive result.
Ulysses had to have an  excessive number of maneuvers due
to a failed nutation damper.  Although the analysis was indicative,
individual errors were as much as an order of magnitude larger than the
effect \cite{pioprd}.} 
The Voyager data is too imprecise to test for the anomaly.\footnote{
The Voyagers are three-axis stabilized.  The resulting often-used gas
jets yield a navigation error of $\sim 10^{-6}$ cm/s$^2$, which is an
order of magnitude larger than the Pioneer anomaly \cite{pioprd}.}
On the other hand, the data is NOT SEEN on large, bound, astronomical
bodies.  

Finally, remember that, although the anomaly is commonly 
{\it interpreted} as an acceleration, the fundamental data 
is a Doppler shift.  The anomaly could be something else, like a time acceleration.    
That being said, What do we only ``suspect'' or do not know?

To begin with, we have no real idea how far out the anomaly goes.   We
only know that it is roughly a constant between 20 and 70 AU, 
that {\it before} Saturn encounter (at around 10 AU) and the transition to
hyperbolic orbit, Pioneer 11 did not show the anomaly, and after 10 AU
Pioneer 10 roughly showed the anomaly. 

An important goal is to perform a good detailed analysis of this early
data. 


\subsection{Lessons learned from studying the Pioneer data}
\label{lessons}

\begin{figure}[h]
\psfig{figure=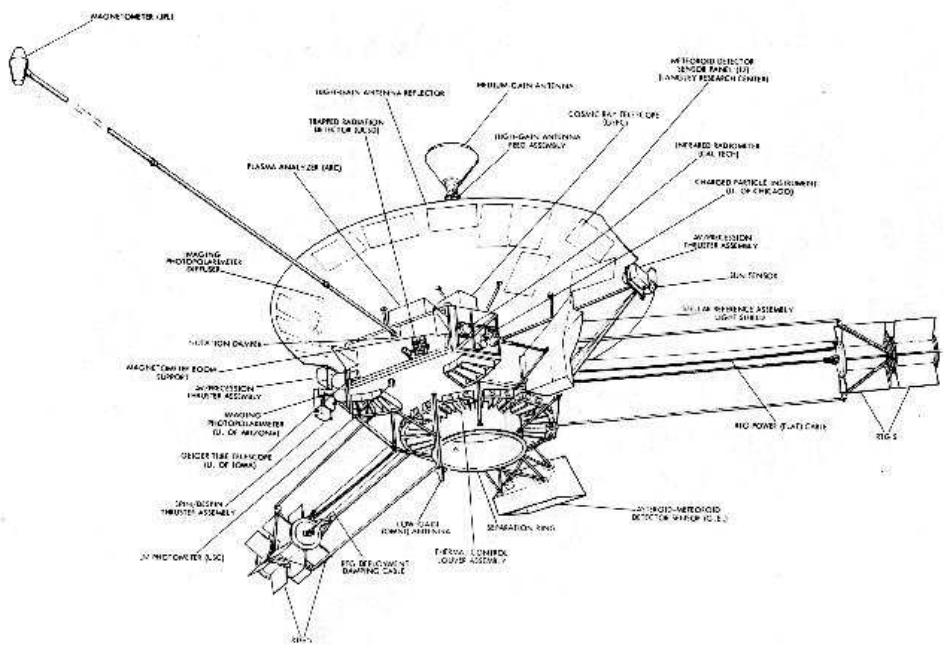,width=5in}
  \caption{A drawing of the Pioneer spacecraft.  
 \label{fig:trusters}}
\end{figure}

The lessons learned from the Pioneers are a guide on how to build a
spacecraft that can accurately investigate the 
anomaly.  Among the most important features of the Pioneer
10/11 spacecraft were their attitude control system, navigation and
communication, on-board power, thermal design, and mission design 
\cite{piomission,piofind,lessons}.  (See Figure \ref{fig:trusters}.)


{\bf Precision attitude control:}
The attitude control system should enable precise acceleration reconstruction
along each spacecraft axis.  This can
be done with spin-stabilized attitude control, as was implemented on the Pioneers.   This is the most
preferable option.  It allows for a minimum number of
attitude correction maneuvers which are, because of the
maneuver-associated propulsive gas leaks, notoriously difficult to
model.  Leakage from thrusters of the propulsion system is the major
navigation problem for 3-axis stabilized vehicles, but its impact is
minimal for spin-stabilized spacecraft.\footnote{If one were to chose 
3-axis stabilization, the uses of accelerometers and/or reaction wheels
would be critical points of concern.}  With spin-stabilization 
the spacecraft spin behavior can also be precisely monitored.  The
understanding of the spin history, coupled with knowledge of all
possible sources of torque, will provide auxiliary information on the
anomaly.  

{\bf On-board propulsion system:} 
For the reasons discussed in the attitude control requirements above,
one would need precisely calibrated thrusters, propellant lines, and fuel
gauges and knowledge of the propellant mass usage history.  However,
currently available sensors may not be accurate enough to yield the
required long-term 3-dimensional acceleration sensitivity.  
 Autonomous real-time
monitoring and control of their performances would also be helpful.  This 
would come, however, at  the cost of an increased telecommunication data
rate.

{\bf Navigation and communication:}
As with the attitude control system, the navigation and communication
system should allow a 3-dimensional acceleration reconstruction at the
level of $\sim 0.01 \times 10^{-8}$ cm/s$^2$ 
for each vector component.  Having both Doppler and range tracking, 
and possibly VLBI and/or $\Delta$DOR, would allow
the precise measurements of plane-of-the-sky angles that are needed
for 3-dimensional acceleration reconstruction.   mrad pointing should
be sufficient to enable precise attitude reconstruction.  The
preferred communications frequencies are X- and Ka-band with
significant dual-band tracking for possible dispersive media
corrections.  Alternatively, optical tracking could be employed.  This
will be further evaluated in the future.  

{\bf On-board power for deep space:}
No deep-space mission can accomplish its goals without a reliable
source of on-board power.  For now, this must be provided by RTGs.
The location of the RTGs is a very critical choice, as it must satisfy
the requirements for inertial balance, stability, and thermal isolation.
For a spin-stabilized option, one would want to position the RTGs as
far away as practical from the bus.  Having the RTGs on extended booms
aids the stability of the craft and also reduces the heat
systematics.  

{\bf Thermal design requirements:}
Thermal design is one of the most critical design issues for a mission
to explore the Pioneer anomalous signal.  
The entire spacecraft and/or probe
should, as much as possible, be heat-balanced and 
heat-symmetric fore/aft.
In particular, the emitted radiant heat from the RTGs 
must be symmetrical in the fore and aft directions and 
the thermal louvers should be placed on the
sides to eliminate fore/aft thermal recoil force due to the release of
excess radiant heat.\footnote{For a 3-axis stabilized craft it would
be harder to balance recoil forces and torques.} 
One should have a precise knowledge of all heat sources -
RTGs, electronics, thrusters, etc.  Also, an active control of
all heat dissipation channels would be an additional critical aid.

Finally, it is important to have a precise knowledge of 
how the spectral properties of the materials, from which the
spacecraft surface is composed, will degrade.  This is a 
challenging issue to discuss quantitatively. 
The difficulty lies firstly in the precise folding of the
reflective insulation blankets and in the precision painting of all the
external surfaces. 
But it is still hard to predict the exact behavior of all the
surfaces on the spacecraft, especially after long
exposure to the space environment (i.e., solar radiation, dust,
planetary fly-byes, etc.). 
Knowing this all would result
in a precise knowledge of the future history of the 3-dimensional
vector of any residual thermal recoil force.

{\bf Modeling external non-gravitational forces:} 
The analysis 
presented in \cite{pioprd} also emphasized the importance of a
precise modeling of the solar radiation pressure.  
For distances below 10 AU, the capability of
measuring, monitoring, and compensating for the effects of solar
radiation pressure may be very important in achieving the precision
radio-science objectives.  In addition, similar information on the
electrical charge accumulated on the spacecraft would be very
important information for the purposes of precise orbit determination
and attitude control.  

{\bf Hyperbolic, solar system escape trajectories:} 
The Pioneer anomaly was found on spacecraft following hyperbolic
escape trajectories at  distances between 20 and 70 AU out from the
Sun \cite{pioprd}.  Although it might have been present closer in,
this has only been imprecisely studied 
\cite{pioprl,pioprd,piomission,piofind}.  For this reason, and also to
reduce the effect of external systematics, the experiment should reach
distances greater than 20 AU from the Sun.  
Achieving small orbit
transfer times to regions further than 20 AU would yield a concept
validation and technology infusion to other missions having a 
demand for rapid access to distant regions of the solar system.
Thus, arriving to this distance from the Sun
in less than 10 years would be most desirable.  

To yield a direct
test for any velocity-dependence in the signal, one also wants the
craft to have a significantly different velocity than the Pioneers.
All this means that when the craft reaches deep space it should be in
a high-velocity, hyperbolic, escape orbit.  As a baseline, the
desirable mission duration should be shorter than 10 years to 20 AU. 

{\bf To summarize:}  We observe that 
the Pioneers were "accidentally" built in a way that yielded very
precise orbits; newer craft will need special designs to surpass the
accuracies of the Pioneers.  Effects that have normally
been considered to be relatively unimportant, such as rejected thermal
radiation, gas leaks, and radio beam reaction, now turn out to be
critical for the precise navigation of science craft in the 21
century.  It is hard not to emphasize the most successful feature and
main Pioneer lesson for a potential spacecraft and mission design to
test the anomalous acceleration - make it simple!   


\section{A Future Test}
\subsection{Mission objectives}

The physics objectives of a mission to test the Pioneer anomaly would be 
two-fold:\footnote{With the growing interest in the anomaly, there have recently been a
number of proposals for missions to test the anomaly 
\cite{piomission}-\cite{pluto}.}

{\bf(1)} to fly a deep-space 
experiment that is capable of achieving a 3D accuracy of 
$\sim 0.01 \times 10^{-8}$ cm/s$^2$ 
for small {\it unmodeled} DC accelerations. This quantity is equivalent to a 
level of $\sim 3 \times 10^{-21}$ s/s$^2$ in clock accelerations. 

{\bf (2)} to determine the direction and physical origin of any anomaly
discovered.

The experiment could be performed on a dedicated mission or else considered for a probe that is jettisoned in deep space from a large mission  (Sec. \ref{addon}). 
 

\subsection{Fore/aft symmetric deep-space mission}

We begin by discussing a fore/aft symmetric spacecraft that can be
considered a baseline concept for any successful mission
\cite{piofind}.  It is designed to perform a general test of 
the anomaly as it was observed. It is also designed to eliminate systematics 
and allow precise tracking from distances in the inner solar system to 
40 AU and beyond. Its design is specifically motivated by the lessons
learned from the Pioneer missions described in Section \ref{lessons}. 

For a nuclear powered spacecraft, the major navigation systematic 
in deep space is thermal emission generated by the spacecraft's power system.  
This is because, with either space-craft centered RTGs or else
a space-craft centered nuclear reactor, 
there are many to hundreds of kW of heat power generated.  This 
produces up to hundreds or more  W of electrical power in the bus.  
The heat dissipation can produce a non-isotropic force on the craft
which can dominate a force the size of the Pioneer anomaly, especially if
the craft is light.   For example,  only $\sim 63$ W of directed power 
could have explained the anomaly the 241 kg Pioneer craft with half
its fuel depleted. 
Therefore, if those RTGs had been placed ``forward''
they obviously would have yielded a huge systematic

The heat systematic will be eliminated by making
the heat dissipation fore/aft symmetric.  In a stroke of serendipitous 
luck, the Pioneer RTGs, with  $\sim$ 2,500 W of heat, were placed at the end of booms.\footnote{Because they were 
the first deep space craft, the Pioneer engineers were worried
about the effects of nuclear radiation on the main bus electronics.
Placing the RTGs at the end of booms was the solution.} 
This meant they had little thermal effect on the
craft. Further, the rotation of the 
Pioneer craft and their RTG fin structure design meant the
radiation was extremely symmetrical fore-aft, with very little heat
radiated in the direction towards the craft. The same concept will be
used for this mission, with perhaps shielding of the RTGs to further
prevent anisotropic heat reflection.   

The electrical power in the equipment and instrument compartments must
also be radiated out so as not to cause an undetected systematic.  For
the Pioneers the central compartment was surrounded by insulation.  
There were louvers forward to be open and 
let out excess heat early in the mission and to be closed and retain
heat later on when the electrical power was less. The electrical power
degrades faster than the radioactive decay because 
{\it any} degradation of the thermoelectric
components means the electric power degrades from this {\it on top of}
the degradation of the input heat due to the 87.74
half-life of the $^{238}$Pu.\footnote{
For the Pioneers, the time from launch to when the Pioneer 10 electrical
power had been reduced to 50\% was about 20 years.\cite{piompla}}
For this mission, the louvers will 
be on the side of the compartment so that they radiate in an axially
symmetric manner as the spacecraft rotates.  

\begin{figure}
        \begin{minipage}
                       {.46\linewidth} 
            \epsfig{file=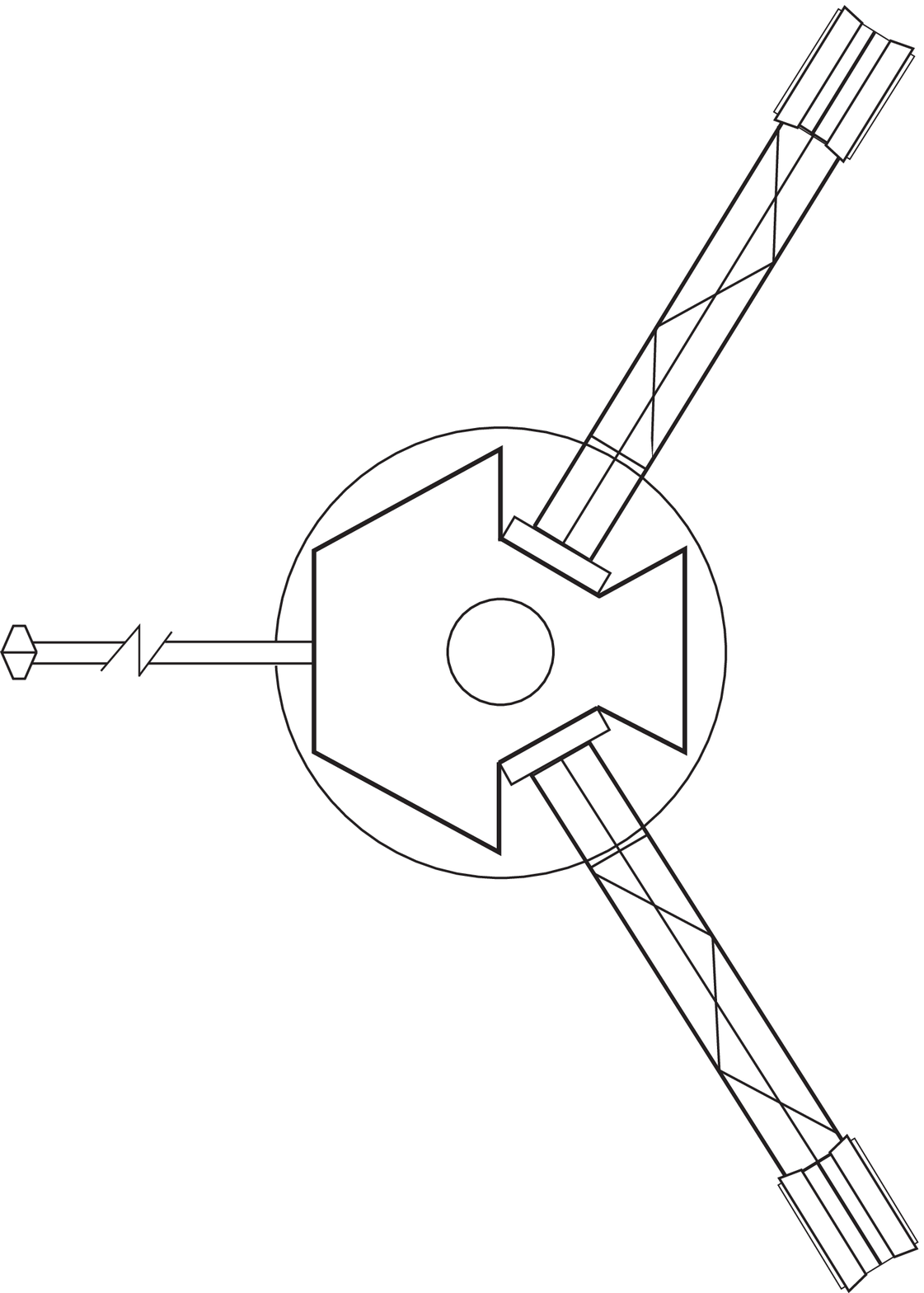,width=65mm}
        \end{minipage}
        \hskip 15pt
        \begin{minipage}
                      {.46\linewidth}
            \epsfig{file=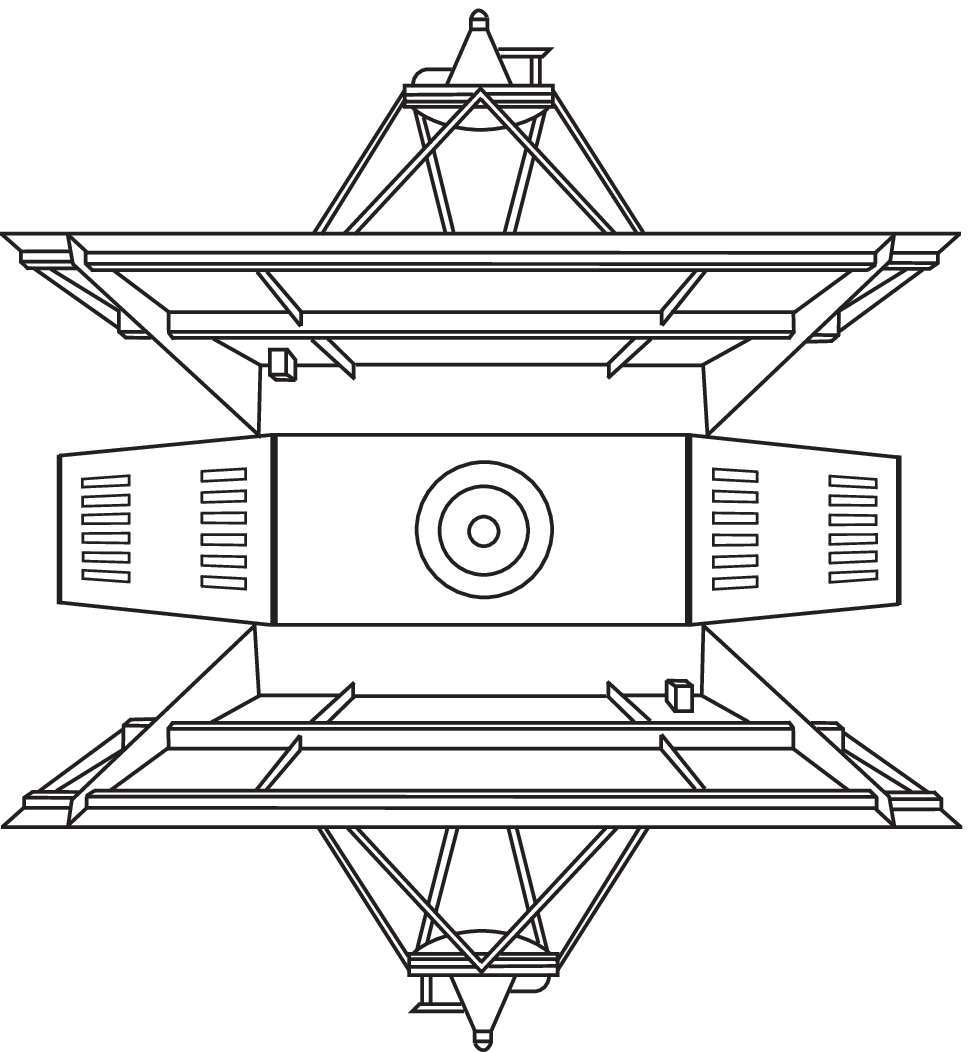,width=65mm}
        \end{minipage}
\caption{The top (left) and side (right) views (different scales)
of ``yo-yo" craft concept \cite{piofind}. 
The scale of the circular antenna is on
the order of 2 to 2.5 m.   The RTGs are deployed on the left.  There
also is an indication of a third long boom where an instrument 
package to detect interstellar matter could be placed. 
Depending on the final mission objectives this instrument
package could be replaced by a third RTG. The side view
shows the louvers radiating to the side and the antennas, taken and
modeled from the Cassini Cassegrain antenna \cite{casantenna}.}
            \label{yoyo}  
\end{figure}

The most unique design feature of this mission is 
the symmetric radio-beam: The dual for/aft symmetric spacecraft design
uses two identical and simultaneously transmitting radio-antennae aligned along the spin-axis and facing in the opposite direction. This is the yo-yo concept described in Ref. \cite{piofind} and shown in Fig. \ref{yoyo}. 
By implementing such a design, one can essentially eliminate the
radio-beam communication bias.  A spacecraft design such as this
has never been proposed before.  


This symmetric antenna design also 
allows us to further minimize the heat systematic.  Any heat
reaching the back of the two antennas. despite insulation placed
in, around, and in the support of the bus, will be reflected
symmetrically fore/aft.
This choice would also eliminate any remaining fore/aft asymmetry 
in the acceleration estimation since, by rotating the craft 
180$^\circ$, one could cancel out any difference in the  two signals.

Thus, this  
fore/aft symmetric design greatly reduces the size of any 
possible heat systematic by its simplicity.\footnote{Recall that for the Pioneers, contributions to the detected anomaly of order  $10^{-8}$ cm/s$^2$ came individually from the RTGs and power dissipation \cite{pioprd,piompla}.} 
Preliminary analysis (see Ref. \cite{piofind}) 
suggests that even with existing technology one could balance the
fore/aft geometry of the spacecraft to minimize the possible
differential heat rejection systematic to the level  
${\leq 0.03 \times 10^{-8}}$ cm/s$^2$.  (With improved technology one
could reach our desired goal.)  

A final factor in the spacecraft heat transfer mechanism  is the
optical properties with time of the spacecraft surfaces.  However, such changes did not seem to affect the Pioneer results despite their rugged voyages \cite{pioprd,piompla}.  Further, this mission's use of
rotating the antennas (described above) will obviate any residual
effect by cancellation.


\subsection{The Direction of the Anomaly}

For the Pioneers at the 20-70 AU studied \cite{pioprd}, the directions 
(1) towards the Sun, (2)
towards the Earth, (3) along the direction of motion of the craft,
or (4) along the spin axis could not easily be distinguished.  
To obtain a determination of the actual direction would be very important, since these directions would tend to
indicate a physical origin that is 
(1) new dynamical physics originating from the Sun, 
(2) a time signal anomaly, 
(3) a drag or inertial effect, or 
(4) an on-board systematic. 

\begin{figure}[h]
\psfig{figure=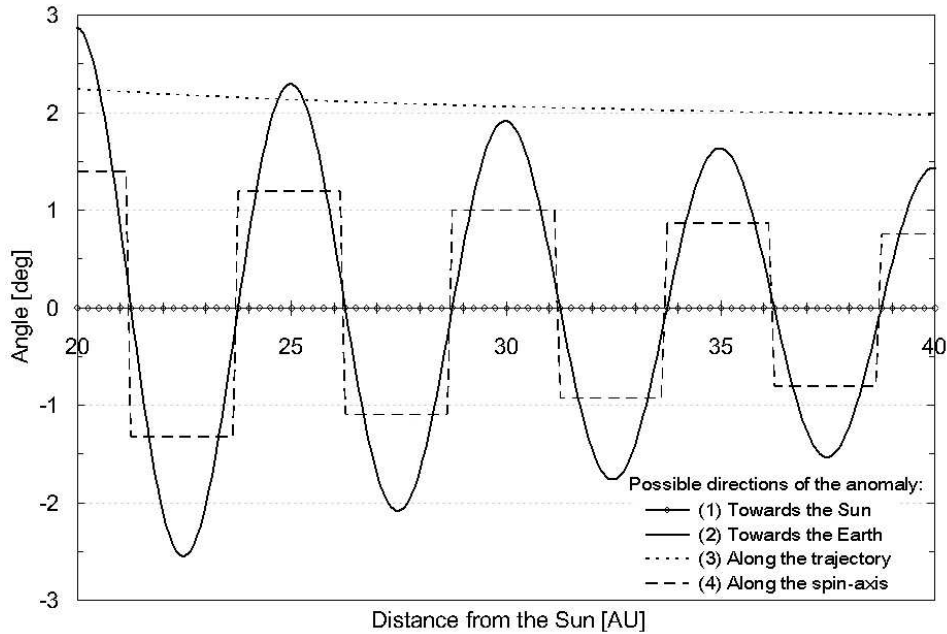,width=5in}
  \caption{The signatures for four possible directions of the
anomalous acceleration acting on the proposed spacecraft. The
signatures are distinctively different and are easily detectable with
the proposed mission. (See the text.) 
 \label{fig:angles}}
\end{figure}

At 20 AU these directions are of order 3 degrees apart (the
maximum angle subtended by the Sun and the Earth (even more
depending on the hyperbolic escape velocity vector).  
In Figure  \ref{fig:angles} we show the angles at which these forces
would act for a hyperbolic trajectory in the ecliptic, between 20 and 40 AU.
The eccentricity is 5 and the craft travels at approximately a terminal 
velocity of 5 AU/yr.  ($a$, the minimum distance from the hyperbola to
its intersecting asymptotes, is 1.56 AU.)  The reference 
curve (1) at zero degrees is the constant direction
towards the Sun.  Other angles are in reference to this. 
Starting to the right in the plane for definiteness, the angle
towards the Earth (2) is a cosine curve which is modified by an $1/r$
envelope as the craft moves further out.  The angle from the Sun
to the trajectory line 
is shown in (3).  Finally, the direction along the spin axis (4) is a
series of decreasing step functions.  This indicates two maneuvers per
year to place the antenna direction between the maximum Earth
direction and the null Sun direction, performed as the Earth passes
from one side of the Sun to the other.


The use of 3-D navigation discussed above
would result in a precise spacecraft positioning with respect to the
solar system barycentric reference frame. As with the Pioneers, the
accuracy of the determination will depend on the properties of the
antenna radiation pattern. Highly pointed radiation patterns are
available for  higher communication frequencies. In order to
be on the safe side, one can use a standard X-band antenna with a
0.5$^\circ$ angular resolution.  Therefore, 
if the anomaly is directed towards the Sun (1), 
the radiometric tracking described above will be able 
to establish such a direction with sufficiently high accuracy. 
 
If the anomaly is directed towards the Earth (2), the current accuracy of
the Earth's ephemerides will be a key to determining this fact.
Furthermore, in this case one would clearly see a dumped
sinusoidal signal that is characteristic to this situation (see
Figure \ref{fig:angles}). The use of standard hardware 
would enable one to accurately establish this direction with a high
signal-to-noise ratio.  

An almost-linear angular change approaching the direction towards the
Sun (also highly correlated with the hyperbolic trajectory) would
indicate a trajectory-related source for the anomaly (3). This situation
will be even more pronounced if the spacecraft were to perform a
planetary fly-by. In the case of a fly-by, a sudden change in the
anomaly's direction will strongly suggest a trajectory-related
source for the anomaly. 

Finally, a step-function-like behavior of the
anomaly, strongly correlated with the maneuver history, would 
clearly support
any anomaly directed along the spin-axis (4).  
As a result, a combination of the standard navigation methods
addressed above in combination of the symmetric spacecraft design
would again enable one to
discriminate between these four different directions of the anomaly
with a sufficiently high accuracy.  

It is clear that these four possible anomaly directions all have
entirely different characters.  The symmetric fore/aft 
mission concept was designed
with this issue in mind. 
In Ref. \cite{piofind} it was  determined that the
 main features of the signatures of 
Fig. \ref{fig:angles} could be distinguished over a year.\footnote{
The use of antennas with highly pointed radiation patterns and of star 
pointing sensors would create even better
conditions for resolving the true direction of the anomaly.}

Thus, this mission will also provide evidence 
on the origin of the anomaly, by helping to determine its direction. 
Determining which of the possible physical causes for the anomaly is the 
correct one will be important 
in the more general frameworks of the solar system
ephemerides as well as spacecraft design and navigation.  


\subsection{Formation flying mission concept}

As these developments have been going on, interest in the anomaly has
spread.  In particular a collaboration has been formed among members of the European science community, which we have joined. 
It has presented a proposal \cite{cosmic} to the European Space Agency (ESA) in answer to its call, Cosmic Vision Themes for 2015-2025.. 
This collaboration is interested in producing a mission that will also develop new technologies, so as to involve the space technology community. 

Formation flying technologies have 
become one of the top priorities for both ESA and NASA.  In
particular, a recent NASA Research Announcement calls for
development of precision formation flying as an enabling technology to meet future high-priority science objectives \cite{nasatech}. 
Due to limited launch
vehicle fairing sizes and the need to phase optical elements
over long distances on flexible structures, separated spacecraft
formation flying is the only viable means to satisfy many demands of
modern science missions.  

Therefore, as a reference design, the collaboration choose to emphasize a
concept that relies on formation flying \cite{enigma}.   
The design has a primary craft that is robust
and able to nurture itself for 7-12 years in the environment of deep
space.  The craft would communicate high-precision Doppler and range
data to Earth, with X- or Ka-band, or perhaps even via optical
communication.  

\begin{figure}[h]
\psfig{figure=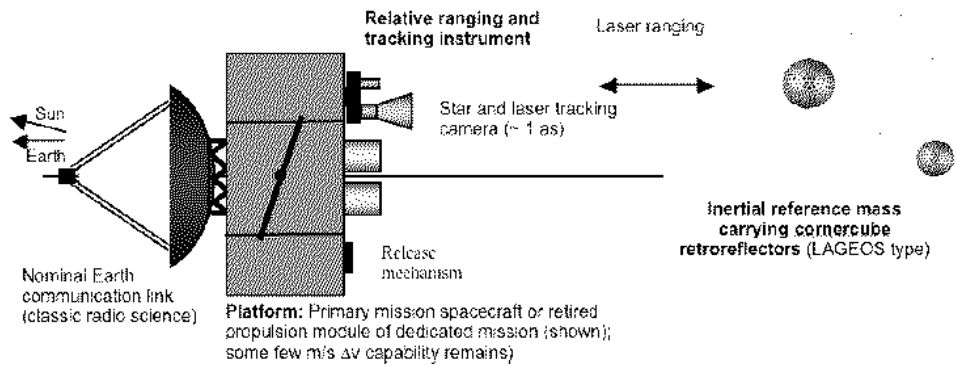,width=5in}
  \caption{Formation flight scenario \cite{enigma}: an active spacecraft (or retired
propulsion module) tracking a locally detached, formation flying
reference inertial masses via ranging and angular sensors.  The
spheres have been previously gently released from the spacecraft.
Additional on-board sensors may include magnetometer, solar radiation
intensity monitors, and a dust spectrometer including both charge and
inertial sensors.  
\label{formation}}
\end{figure}

The idea is to avoid the inherent problems of self disturbance
of an inertial sensor on board the primary spacecraft by placing the
inertial reference mass(es) (i.e. subsatellite(s)) outside the craft at a
sufficient, but not too large distance.  A laser ranging sensor that
employs a mW laser monitors the 3-dimensional vector of mutual
separation between the spacecraft and subsatellite.  Any 
subsatellite is covered with corner-cube retro-reflectors that
enable precise laser ranging similar to that currently used for satellite and
lunar laser ranging.  Figure \ref{formation} shows a generic
concept \cite{enigma}.\footnote{A related, but more complicated,
idea has also been proposed \cite{chui}.}


To enable this concept three requirements must be fulfilled:  i) the
subsatellite must be at a sufficient distance from the main craft so
that any radiant heat from the primary craft will not affect the
motion of the subsatellite; ii) the reflected laser light must also
not cause a significant force on the subsatellite; iii) the primary
craft must be able to laser range to the subsatellite to determine its
relative position history; i.e., 
the primary craft must be able to ``follow" the subsatellite.  
The objective of this architecture is to precisely measure the motion
of the inertial reference mass and to eliminate the effect of any
disturbance of the primary spacecraft on the subsatellite. 

Conceptually a combination of two observables related to the precise
distance determination between the Earth and the spacecraft and with
that between the spacecraft and the subsatellite would yield a more
accurate distance between the Earth and the subsatellite.
In principle the spacecraft does not have to be dynamically quiet as long as the
relative distance and lateral position remain within the laser
tracking capability.\footnote{Current estimates suggest that tolerable
distances are up to few km with a lateral imbalance of up to 5 degrees.}

The relative
range, range rate and lateral tracking is a 3-dimensional two-step
process and needs to be sufficiently accurate in relative angle
determination between main link to Earth and local link to reference.
Despite the advantages, the mission operation has some added
complexity.  Also, the error in the two-step measurement process has not  been studied for a noisy main craft.  This needs to be done to be sure the concept works.   

The spacecraft has to provide an accurate  release mechanism for the reference masses.  Depending on its stabilization mechanism, its other properties could be similar to those of the fore/aft symmetric mission; it would not need to perform many large  maneuvers, the dedicated payload, including reference masses, could be in the range of 150-250 kg, it would use the only source of autonomous power in deep space, RTGs. It would need at least about 50-75 W of electrical power to run
the communications and heat the fuel (for maneuvering).  Since the
electrical thermocouples degrade with time, the power at launch should
be of order 140 W.


\subsection{Propulsion Options}

The launch vehicle is a major consideration for any deep-space
mission.
Propulsion systems are quite literally the driving force behind any 
effort to get a payload into space, especially on the interplanetary 
orbit.  

To test the Pioneer anomalous acceleration in the most suitable 
environment, one wants to reach a distance greater than 15 AU from the 
Sun. In this region one can clearly distinguish any effect from solar 
radiation pressure, interplanetary magnetic fields, as well as solar and 
interplanetary plasmas. A fast transfer orbit is very desirable, to 
allow reaching the target region in a reasonable time. Therefore, a large 
solar system escape velocity is desired (say, more than 5-10 AU/yr). In 
contrast, the Pioneers and Voyagers are cruising at velocities 
of about 2 and 3 AU/yr. One needs something faster than that.

The obvious first idea is a very energetic chemical 
propulsion rocket. An escape terminal velocity of $\sim$ 5 AU/yr is
achievable with current mission design technologies and   
existing heavy launch vehicles (Ariane 5, Proton, Delta IV, or Atlas V)
\cite{piofind,rockets}. 

However, it must be noted that purely chemical propulsion,  
even using present day powerful and expensive (!) launchers,  
cannot achieve the desired short mission times without 
a powerful chemical kick stage and/or suitable planetary gravity-assist 
maneuvers. That immediately sets severe constraints on any mission 
profile.   On the other hand, if a chemical launch vehicle were to be
used for a dedicated mission,  
with gravity-assist flybys, a mission could also address the
question of if the Pioneer 11 anomaly started near its Saturn flyby,
when it reached escape velocity. 

Further, the use of chemical propulsion is at the limit of its 
capabilities to satisfy the needs for deep-space exploration. For this 
reason, both ESA and NASA have initiated programs to study alternative 
propulsion methods for their deep space exploration missions. 
For this reason, the new collaboration also decided to consider
various new technologies that might be employed.  They include  
1) nuclear electric, 2) solar electric, 
and 3) solar thermal propulsion.  

But the choice with most widespread interest in Europe is solar
sails.  The Europeans have had a long interest in this idea, 
as exemplified by the ESA/German Odissee concept 
\cite{odissee}.   With ESA's \cite{odissee,dach} and NASA's 
\cite{garner2000,liewer2002} 
separate interest in this solar sail  technology, it may prove useful and
also could yield science \cite{piosail}.

The ultimate hope is to obtain sail materials with a weight of 1 g per m$^2$ 
of sail area and light weight structures to support a sail of size 200 m in 
length.  For now this goal is in the future, but the advent of multi-cm long
nanotubes may be pointing the way.  With such a sail, one can envision 
accelerating to velocities over 14 AU/yr by the time the orbit of Jupiter is 
reached, jettisoning the sail, and having the main mission coast.   


\subsection{Testing the anomaly as part of a larger mission}
\label{addon}

Given that any mission to very deep space will need a large terminal velocity and the concomitant commitment of effort and funds, the question naturally arises of if a test of the Pioneer anomaly can be done, not as a dedicated mission but rather, as a probe on a very large mission that would be jettisoned to fly on separately after the requisite distance and speed had been obtained \cite{piomission,piofind,pluto}.  In principle 
this could happen with a mission that used nuclear reactor power.    

As an example, it might be considered for an add-on to a mission that would use a reactor-ion engine, such as Prometheus \cite{prometheus}.  Indeed, this idea has already been discussed  in the context of  the InterStellar Probe mission to the heliopause \cite{funsten}.   Such a marriage may prove to be an interesting possibility for a test of the Pioneer Anomaly. 



\begin{theacknowledgments}
We all are thankful for the many contributions of the original Pioneer collaboration (Sec. \ref{orig}) as well as those from the new collaboration \cite{cosmic}. 
M.M.N.  thanks Alfredo Mac\'ias and Claus L\"ammerzahl for their kind hospitality and acknowledges support  by the U.S. DOE.
The work of S.G.T  and J.D.A. was performed at the
Jet Propulsion Laboratory, California Institute of Technology, under
contract with the  National Aeronautics and Space 
Administration.
\end{theacknowledgments}


\hyphenation{Post-Script Sprin-ger}


\end{document}